\documentclass[5p]{elsarticle}

\usepackage[T1]{fontenc}
\biboptions{longnamesfirst,sort&compress}

\usepackage{url}
\usepackage{siunitx}
\usepackage[]{graphicx}
\usepackage{booktabs}
\usepackage{multirow}
\usepackage{amsmath}
\usepackage{amsfonts}
\usepackage{bbold}
\usepackage{tipa}
\usepackage{bibentry}
\usepackage{hyperref}
\usepackage{rotating}
\usepackage{subcaption}
\usepackage{makecell}

\usepackage{enumitem}
\setlist[enumerate]{itemsep=0pt}
\setlist[itemize]{itemsep=0pt}

\usepackage{array}
\newcolumntype{L}[1]{>{\raggedright\let\newline\\\arraybackslash\hspace{0pt}}m{#1}}
\newcolumntype{C}[1]{>{\centering\let\newline\\\arraybackslash\hspace{0pt}}m{#1}}
\newcolumntype{R}[1]{>{\raggedleft\let\newline\\\arraybackslash\hspace{0pt}}m{#1}}

\graphicspath{{figures/}}

\newcommand{\mtrx}[1]{\ensuremath{\MakeUppercase{\boldsymbol{#1}}}}

\begin{document}

\begin{frontmatter}

\title{Investigation of the Assessment of Infant Vocalizations by Laypersons}

\author[labp,fimn]{Franz Anders \corref{mycorrespondingauthor}}
\ead{franz.anders@htwk-leipzig.de}

\author[fimn]{Mario Hlawitschka}
\cortext[mycorrespondingauthor]{Corresponding author}
\ead{mario.hlawitschka@htwk-leipzig.de}

\author[labp]{Mirco Fuchs}
\ead{mirco.fuchs@htwk-leipzig.de}

\address[labp]{Laboratory for Biosignal Processing, Eilenburger Stra\ss{}e 13, 04317 Leipzig, Germany}
\address[fimn]{Faculty of Computer Science and Media, Gustav-Freytag-Stra\ss{}e 42a 04277 Leipzig, Germany}
\address{Leipzig University of Applied Sciences, Germany}

\begin{abstract}
The goal of this investigation was the assessment of acoustic infant vocalizations by laypersons. More specifically, the goal was to identify (1) the set of most salient classes for infant vocalizations, (2) their relationship to each other and to affective ratings, and (3) proposals for classification schemes based on these labels and relationships. The assessment behavior of laypersons has not yet been investigated, as current infant vocalization classification schemes have been aimed at professional and scientific applications. The study methodology was based on the Nijmegen protocol, in which participants rated vocalization recordings regarding acoustic class labels, and continuous affective scales valence, tense arousal and energetic arousal. We determined consensus stimuli ratings as well as stimuli similarities based on participant ratings. Our main findings are: (1) we identified 9 salient labels, (2) valence has the overall greatest association to label ratings, (3) there is a strong association between label and valence ratings in the negative valence space, but low association for neutral labels, and (4) stimuli separability is highest when grouping labels into 3 -- 5 classes. We finally propose two classification schemes based on these findings.

\end{abstract}


\end{frontmatter}

\section{Introduction}
\label{sec:introduction}

Infants you their voice as on of their primary communication tools: They express distress through crying, joy through laughing, and babble in development of the speech capacity. Consequently, listeners regularly assess infant vocal expressions for a variety of assessment goals. \citet{stark1981infant}s

Untrained parents usually asses infant vocalizations intuitively and subjectively. However, in scientific or professional applications we require \emph{reliability}. This means that the evaluation of vocal expression should be consistent withing and across observers. We achieve this through \emph{assessment tools}. These usually comprise of two parts: (1) A \textbf{classification scheme}, which is a listing of the relevant infant vocalization classes, and (2) an instruction which associates occurrence of such classes with meanings. There are also catalogue-like classification schemes which list classes for their own sake, e.g. so that they might be referenced in other assessment tools. 

However, currently there is no single, unified classification scheme for infant vocalizations. One of the main reasons is there are many assessment tools for different assessment goals which all define their own classification schemes.

\citet{buder2013acoustic} presented one of the most popular catalogue-like classification schemes. It differentiates between so called \emph{reflexive sounds} and \emph{protophones}. Reflexive sounds include: distress vocalizations (e.g. cries and fusses), laughs, and vegetative sounds (breathing, burping, coughing, sucking etc.). On the other hand, protophones are volitional, and related to the acquisition of the speech capacity. Examples for protophones are vowels, quasi-vowels, squeals, raspberries, etc. Protophones also involve sequences of repeated sounds, such as marginal or canonical babbling (which correspond to what the layperson usually imagines as stereotypical babbling).

Various assessment tools for vocal development refer to this catalogue. For example, the SAEVD-R by \citet{nathani2006assessing} associates infant vocalization classes  the expected ages of emergence. To assess vocal development, we monitor whether these vocalization classes emerge at the appropriate age. There are various alternative assessment tools for vocal development, e.g. see \citet{oller2000emergence} and \citet{stark1981infant}. 

\emph{Pain scales} are assessment tools for infant pain in pediatrics. They typically define classes for negative affective vocalizations which map to pain scores \cite{mcgrath2013oxford}. For example, the pain scale NIPS defines the classes \& pain scores \emph{no crying}$=0$; \emph{whimper, mild moaning} $=1$; \emph{vigorous cry, loud scream} $=2$ \cite{lawrence1993development}. Contrary to assessment tools for vocal development, they usually put all neutral and positive vocalizations into the same class. For further examples of pain scales, see \cite{krechel1995cries, merkel1997flacc, hummel2003n}. 

Then, there are classification schemes developed specifically by scientists for their respective studies. \citet{scheiner2002acoustic} investigated developmental and emotional changes in infant vocalizations. For this, they defined a classification scheme of 12 call types. \citet{lin2009infants} studied the changes in infant expressive behavior to mothers, by assessing occurrences of negative, neutral and positive vocalizations.

However, all of these assessment tools and classification schemes have common denominators: (1) They were developed for instructed professionals, and require extensive training to be applied correctly. \citet{nathani2001beyond} published a paper where they elaborated on difficulties of coding infant vocalizations, and the amount of training required for doing so reliably. (2) Classification schemes are  independent of one another, except for those which state explicit references. Even if various schemes define the class \emph{crying}, the term might be defined different in each one. For example, \cite[chapter 11]{barr2000crying} listed over 20 coexisting definitions of \emph{crying} in the year 2000.

There is little research on the assessment of infant vocalizations by laypersons. Studies which employed laypersons for vocalizations assessment did so primarily for rating of distress vocalizations (i.e. fuss- and cry vocalizations). \citet{xie1996automatic} asked parents to rate the level-of-distress of cry-like signals on a continuous scale. The goal was to identify correlations between the level-of-distress and acoustic properties. \citet{barr1988parental} asked parents to document their infants crying behavior in dictionaries. They were asked to discriminate \emph{content}, \emph{fussing} and \emph{crying}. 

Studying the assessment behavior of laypersons for infant vocalizations is of importance for various reasons. First, to understand which classes are reliably and intuitively discriminated by laypersons in domestic environments. Second, it is the basis for labeling large data sets for training machine learning algorithms, as crowd-sourcing data set labeling is the most efficient way for data labeling. Third, to increase knowledge on the universally recognized infant vocalization classes, prior to training in vocalization coding.

The central goal of this study was the development of an classification scheme for infant vocalizations based on the discrimination capability and assessment behavior of laypersons. The precise research questions were:

\begin{itemize}
\item What are the salient labels used for naming infant vocalizations?

\item What is the relationship between those labels?

\item How do labels relate to affect assessment, such as valence?

\item Which classification schemes can be derived based on these salient labels and their relationships?
\end{itemize}

The methodology for this study followed the \emph{Nijmegen Protocol} \cite{slobin2014manners}, a method for uncovering linguistically expressed classes and taxonomies. This involved a survey which presented participants with a set of acoustic stimuli, which they rated regarding textual labels and continuous affect scales. We derived the salient labels and their affect ranges by aggregating ratings across participants for each stimulus. Additionally, we investigated the relationship between labels by analyzing intra-rater similarities of stimuli based on label and affect ratings. Finally, we derived taxonomies as semantic maps based on the discovered salient labels and their relationships.

\section{Materials and methods}
\label{sec:voctax_methods}

The structure of this methods section is as follows: Section \ref{sec:voctax_method_nijmegen} introduces the reference approach this study was based on. Section \ref{sec:voctax_method_acoustic_database} presents the acoustic data set. Section \ref{sec:voctax_method_procedure} presents the survey procedure. Section \ref{sec:voctax_method_rating_items} presents the survey rating items in detail. Section \ref{sec:voctax_method_participants} summarizes the participants of the study. Finally, section \ref{ref:voctax_method_analysis} presents the analysis methods applied to the participant ratings.

\subsection{Study framework: Nijmegen protocol}
\label{sec:voctax_method_nijmegen}

The study followed a method which is referred to as the \textbf{Nijmegen protocol}. This is a survey-based approach for the discovery of linguistically expressed classes inside a domain. \citet{majid2008cross} first applied this method to identify universally recognized classes of material destruction, such as \emph{cutting} and \emph{breaking}. \citet{slobin2014manners} refined the method and proposed the term \emph{Nijmegen Protocol}. They applied it to identify classes for manners of human gait, such as \emph{running}  or \emph{crawling}. \citet{anikin2018human} applied it to auditory assessment of human non-linguistic vocalizations and determined four basic call types \emph{laugh}, \emph{cry}, \emph{scream} and \emph{moan}. That study primarily inspired the application of the framework here, as it shares similar goals, however applied to infant vocalizations instead of vocalizations from adults. We summarize the method as follows:

\begin{enumerate}

\item \textbf{Stimuli selection:} First, a set of stimuli is selected. They should be representative for the variability inside the investigated domain.

\item \textbf{Rating:} Participants rate stimuli in a survey. In this method, the primary rating items are textual descriptions, i.e. labels. Some studies let participants provide free textual descriptions \cite{slobin2014manners}, while other studies used predefined, finite label sets \cite{anikin2018human}. Some studies included additional rating items, such as affect assessments \cite{anikin2018human}.

\item \textbf{Calculation of stimuli similarities:} For each pair-wise combination of stimuli a similarity score is determined. A similarity score is calculated for each participant individually and then averaged across participants. This means that similarities measure the average intra-rater similarity, \emph{not} the inter-rater similarity. This allows participants to speak different languages and consequently discover cross-cultural classifications. The resulting distance matrix is then applied to unsupervised learning algorithms, such as clustering or dimensionality reduction algorithms. \cite{slobin2014manners, anikin2018human}
\end{enumerate}

\subsection{Acoustic data set}
\label{sec:voctax_method_acoustic_database}

Extraction of vocalizations as segments for assessment from recordings requires a segmentation criterion. Most previous studies extracted vocalizations as \emph{utterances}, i.e. the segmentation criterion were the pause between vocalizations. A similar segmentation criterion is the \emph{breath group}, in which a vocalization segment corresponds to a breath-cycle. \cite{scheiner2002acoustic,nathani2006assessing,green2011screaming,buder2013acoustic,oller2013functional, anikin2018human, nathani2001beyond} In the framework of \citet{kershenbaum2016acoustic}, these studies operated on vocalizations as \emph{units}.

However, particularly studies on perception of cry-like infant vocalizations showed that human listeners incorporate acoustic information spanning various units, such as the rhythm of units \cite{xie1996automatic,wood2009changes,zeifman2004acoustic,gustafsson2013fathers, barr2000crying}. We additionally hypothesized that laypersons would be me comfortable by judging short collections of units, rather than units in isolation. 

Therefore, we aimed stimuli to be recording excerpts of approximately \SI{6}{\second} spanning various utterances. In the framework of \citet{kershenbaum2016acoustic}, we operated on \emph{sequences}, as opposed to \emph{units}. However, if a unit happened to be present in isolation in a recording, i.e. surrounded by a long pause, it was also considered, however including some of that pause.

We collected audio recordings of infant vocalizations from freely available online sound libraries. The sources were: \url{https://freesound.org/}, \url{http://www.bigsoundbank.com/}, \url{http://www.soundarchive.online/}, \url{https://www.zapsplat.com/}, \url{http://soundbible.com/} and \url{https://www.soundjay.com/}. All recordings using the keywords \emph{baby} or \emph{infant} were considered. We discarded recordings which met at least one of the following criteria:

\begin{enumerate}
\item Insufficient recording quality: This included constant and high amounts of background noise or sounds, excessive reverberation and echo, soundeffects etc.
\item Shorter than \SI{4}{\second}: These recordings provided insufficient temporal context according to the discussion above.
\item  Infants older than 9 months: Infant vocalizations increase in complexity with age. According to the SAEVD-R~\cite{nathani2006assessing} infants enter the vocalization stage \emph{advanced forms} at 9 months. We excluded this stage to limit the complexity of the \emph{babbling}-class. The age was determined through the description of the online source. If the description was absent, we determined acoustically whether a signal contained advanced form babbling as described in~\cite{nathani2006assessing} and excluded such signals.
\end{enumerate}

The resulting database consisted of 228 sound files with a total duration of $\SI{7910}{\second}$ and a mean duration of $\SI{34}{\second}$. We extracted segments from these recordings as stimuli candidates, aimed at a length of $4$ -- $\SI{8}{\second}$ according to the discussion above. Segments had to contain at least one vocalization, even if it was just slight breathing, i.e. We did not extract segments with complete absence of vocal activity. We allocated segments to contain units which were acoustically consistent. Units were contained completely in segments, i.e. not cut across. 

The number of resulting segments was 883, with a mean duration and standard deviation of $\SI{6}{\second} \pm \SI{0.8}{\second}$, and a total duration of $\SI{5217}{\second}$. The set of segments displayed a wide variety of vocalizations, ranging from stereotypical vocalizations such as babbling or crying to barely audible vocal activity, such as quite breathing.

The pool of segments had to be reduced to a reasonable amount for participant rating. To ensure that the drawn sample displayed sufficient acoustic diversity, we evaluated each segment regarding the affect dimensions \emph{valence} and \emph{arousal} as defined by \citet{russell1980circumplex} with continuous scales. We clustered  segments based on these ratings through the k-means algorithm \cite{arthur2007k} with 10 clusters. We randomly sampled 10 segments from each cluster.

The resulting set contained 100 segments, which were the stimuli used for this study.

\subsection{Procedure}
\label{sec:voctax_method_procedure}

The survey was provided through a customly programmed web site. Participants conducted the survey at home.

\begin{figure}[htbp]
    \centering
    \includegraphics[width=0.45\textwidth]{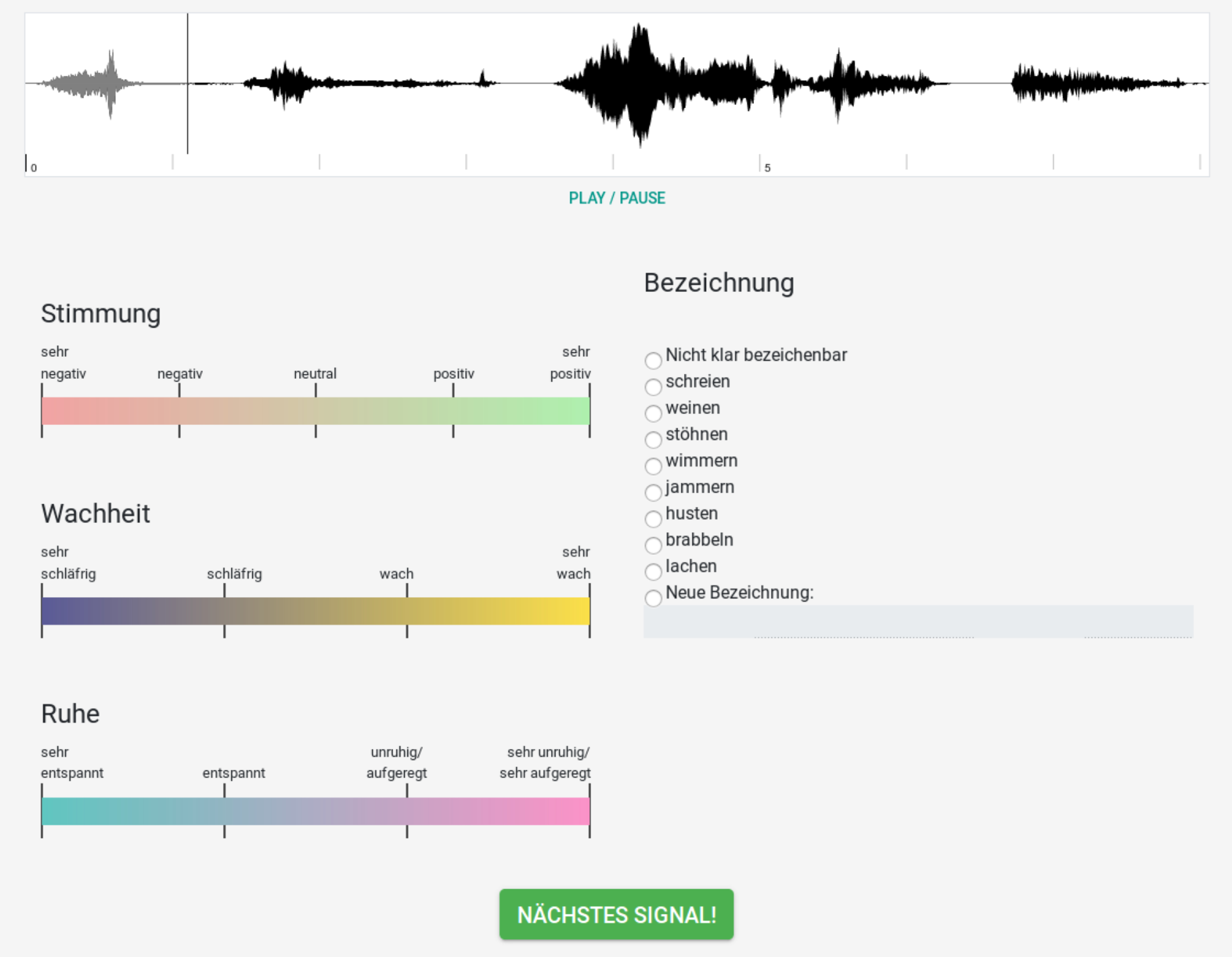}
    \caption[Study (A): Survey main window for rating of an recording]{\textbf{Survey main window for stimulus rating.} Details are provided in the text.}
    
    \label{fig:voctax_method_survey_mainwindow}
\end{figure}

Figure \ref{fig:voctax_method_survey_mainwindow} shows the survey main window for stimulus rating. Participants were presented with one signal at a time. The top part presented a stimuli's wave form with a progression marker, similar to digital audio workstations. Participants could play and repeat the signal autonomously through buttons. The middle panel contained the rating items: Three continuous rating scales \textbf{Stimmung} (\textbf{valence}), \textbf{Wachheit} (\textbf{energetic arousal}) and \textbf{Ruhe} (\textbf{tense arousal}); and one list \textbf{Bezeichnung} (\textbf{label}) for providing a textual description. Section \ref{sec:voctax_method_rating_items} describes the rating items in detail. The bottom panel contained a button for proceeding to the next signal. There was no time limit for item rating. Additionally, participants could save, quit and resume rating sessions, i.e. they could distribute completing the survey over various sessions.

The first page of the survey presented introductory information. This comprised the survey goal, survey procedure and usage of the rating items. \ref{sec:appendix_voctax_introtext} states the instructions on the rating items. The second page listed five signals with rating items for participants to get used to the rating procedure. Participants were informed that answers to these signals were discarded. The warm up signals were preselected to cover diverse and stereotypical infant vocalizations, e.g. stereotypical crying and babbling.

At the end of the survey, participants were asked about their personal data: gender (female or male); age (5-year intervals of 15 -- 19, 20 -- 24, $\dots$); whether they were parents; whether they had professional working experience with infants such as baby sitting. For each of the four rating scales, participants should indicate how challenging they experienced their usage on a scale from 1 -- 5, labeled with \emph{very easy}, \emph{easy}, \emph{modest}, \emph{hard} and \emph{very hard}.

\subsection{Rating items}
\label{sec:voctax_method_rating_items}

\subsubsection{Label}
\label{sec:voctax_method_rating_label}

The rating item \emph{label} asked for a textual description of the vocalization, i.e. naming the acoustic category. We provided a starting pool of labels, which could be expanded freely through custom labels.

The starting pool was as follows (English translations are provided in parenthesis): schreien (screaming); weinen (crying); jammern (whining or fussing); wimmern (whimpering); st\"ohnen (moaning), brabbeln (babbling); lachen (laughing); husten (coughing); nicht klar bezeichenbar (no clear label)\footnote{The stated English translations are meant to be literal translations, not semantically equivalent translations. For example, it might be argued that \emph{crying} and \emph{schreien} are closer semantically.}. The former five terms originate from the only validated German pain scales KUSS \cite[chapter~A3]{zernikow2013schmerztherapie} and BPSN \cite{cignacco2004pain}. The terms \emph{babbling}, \emph{laughing}, and \emph{coughing} were determined in a pilot study, were all participants entered these terms. The option \emph{no clear label} was provided for cases where participants felt that none of the starting labels applied, but neither could think of a fitting custom label. The introductory text encouraged raters to rather think of new labels than use this option. 

Response options were provided via a radio button list. There was no preset value to encourage deliberate choices. Only one term could be chosen. Participants were instructed to select the more salient label if they thought to hear the presence of more than one. 

New labels were added through a text field. Newly added labels were permanently added to the respective participant's label list, so they could be accessed later in the study analogous to the starting pool labels. Added labels were restricted to verbs consisting of a single word, i.e. words with the suffix \emph{ing} (German: \emph{en}) without spaces nor special characters. We imposed this restriction to promote monolexemic, elementary call types, rather than complex descriptions of the situation, similar to the study of \citet{anikin2018human}.

The introductory text highlighted that starting pool labels did not have to be used exhaustively. Participants could ignore starting pool labels which they felt were redundant or not present throughout the survey. 

The final summary page asked participants to indicate whether the starting pools contained \textbf{synonym pairs}, i.e. labels which were redundant or particularly hard to differentiate throughout the study. We indicated that these labels would be mapped together for the respective participants.

\subsubsection{Affect dimensions}

The remaining rating items were three continuous scales for rating of the estimated infants' affect. We adapted the affect model of \citet{schimmack2000dimensional, schimmack2002experiencing} which comprises the following scales: Valence (also known as pleasure) ranging from pleasant to unpleasant, energetic arousal ranging from sleepy to awake, and tense arousal ranging from calm to nervous. We chose to include affect ratings additional to the labels for the following reasons:

\begin{itemize}
\item Other studies which employed laypersons for rating of infant vocalizations most commonly required continuous ratings, rather than categorical descriptions. These studies were limited to rating of cry-like vocalizations, and used the dimension \emph{level-of-distress} \cite{wood2009changes, xie1996automatic}. However, our study encompassed the entire domain of infant vocalizations rather than just cry-like vocalizations. Therefore, we employed a general affect model. We first tested the model by \citet{russell1980circumplex} in a pilot study, containing just two dimensions valence and arousal. However, participants raised concerns about the arousal dimension being ambiguous. Therefore we split the arousal dimension into tense and energetic arousal as proposed by \citet{schimmack2000dimensional} \cite{schimmack2002experiencing}.

\item Any correlations between affect and label ratings are exploitable for designing classification schemes, as they indicate another level of stimuli similarity which might not be obvious from label ratings alone. Particularly cry-like vocalizations have repeatedly been shown to be a \emph{graded signal} \cite[chapter 2]{barr2000crying}. In the most extreme case, affect ratings and labels could turn out to be completely redundant. \citet{anikin2018human} discovered a strong but not perfect association between call types and affect ratings in a similar study.
\end{itemize}

We implemented affect ratings as continuous scales (see Fig. \ref{fig:voctax_method_survey_mainwindow}) with 101 ticks. 

\begin{itemize}
\item \textbf{Valence}, labeled \emph{Stimmung} had a range of $[-50, 50]$, with highlight labels \emph{sehr negativ} (very negative), \emph{negative}, \emph{neutral}, \emph{positiv}, \emph{sehr positiv} (very positive) at ticks $-50, -25, 0, 25, 50$ respectively.
\item \textbf{Energetic arousal} labeled \emph{Wachheit} had a range of $[0, 100]$ with highlight labels \emph{sehr schl\"afrig} (very sleepy), \emph{schl\"afrig}  (sleepy), \emph{wach} (wake), \emph{sehr wach} (very wake) at ticks $0,33,66, 100$ respectively.
\item \textbf{Tense arousal}, labeled \emph{Ruhe}  had a range of $[0, 100]$ with highlight labels \emph{sehr entspannt} (very relaxed), \emph{entspannt} (relaxed), \emph{unruhig/aufgeregt} (restless/excited), \emph{sehr unruhig/sehr aufgeregt} (very restless/excited) at ticks $0,33,66,100$ respectively.
\end{itemize}

The introductory text highlighted that scale ticks were only for orientation, and that values could be chosen continuously. There were no preset values to encourage deliberate value choices.

\subsection{Participants}
\label{sec:voctax_method_participants}

A total of 23 persons participated. The gender distribution was 12 male and 11 female. The age range was 20 -- 59, with a median age of 25 -- 29. 7 of the participants were parents, and 4 had professional working experience with infants. All participants completed the survey from start to finish.

\subsection{Analysis methods}
\label{ref:voctax_method_analysis}

\subsubsection{Post-processing of label ratings}
\label{sec:voctax_method_labelpost}

Postprocessing meant manipulation of participant ratings after survey data collection was completed in preparation for the analysis. We applied the following steps for the label ratings:
\begin{itemize}
\item We fixed obvious spelling mistakes of newly added labels. In some cases, this resulted in labels which originally were different across raters to be mapped to the same label.
\item If participants indicated \emph{synonym pairs} at the end of the study, we unified those labels for each participant to the one she or he used more frequently in the survey.
\end{itemize}

\subsubsection{Measurement of stimuli distances}
\label{sec:voctax_method_distance_calculation}

We represented participant ratings of stimuli through matrices $\mtrx{R}^{(L)}, \mtrx{R}^{(V)}, \mtrx{R}^{(E)}, \mtrx{R}^{(T)}$, where the upperscript $(I)$ indicate rating items (e.g. $L = $ label, $V = $ valence, $E = $ energetic arousal, $T = $ tense arousal). $R_{p,i}$ is the rating of participant $p \in \{1,\dots,P\}$ for item $i \in \{1,\dots, 100\}$, where $P$ is the amount of participants. Based on these we calculated distance matrices for stimuli $\mtrx{D}^{(L)}, \mtrx{D}^{(V)}, \mtrx{D}^{(E)}, \mtrx{D}^{(T)}$, where $D^{(I)}_{i,j} \in [0,1]$ indicates the distance between stimuli $i$ and $j$ for the respective rating item. Consequently, $\mtrx{D} \in [0,1]^{100 \times 100}$.

The label based distance $\mtrx{D}^{(L)}$ was calculated as
\begin{equation}
D^{(L)}_{i,j} = \frac{1}{P} \sum_{p = 1}^P I(R^{(L)}_{p,i}, R^{(L)}_{p,j}),
\end{equation}

where $I$ is the indicator function. Consequently, this distance measures the percentage of participants which \emph{individually} provided the same label.

For the affect rating items we calculated distances as

\begin{equation}
D^{(C)}_{i,j} = \frac{1}{P} \sum_{p = 1}^P \frac{| R^{(C)}_{p,i} - R^{(C)}_{p,j} | }{S_{p,C}},
\end{equation}

where $C \in$ \{valence, energetic arousal, tense arousal\} is the scale index, and $S_{p,C} = \max_p(R^{(L)}_{p,i}) - \min_p(R^{(L)}_{p,i})$ is the range for scale $C$ of participant $p$.

When combining various items into distance calculation, we summed item-based matrices.

In the context of the Nijmegen protocol, this calculation of the label-based distance corresponds to the established approach of counting equal descriptions \cite{slobin2014manners, anikin2018human}. In the context of unsupervised machine laerning, the approach corresponds to calculating distances through the Gower coefficient \cite{gower1971general}. This distance metric is specialized in mixed-type variables, where nominal variables (label) employ the ordinary discrete metric, and continuous variables (affect scales) employ the range-normalized L1-distance. Each rater and item is treated as a unique feature.

\subsubsection{Aggregation of stimuli ratings across raters}
\label{sec:voctax_method_rating_agreggation}

One step in analysis of results was determining a single \textbf{consensus rating} for each stimulus and rating item. To accomplish this, we aggregated ratings across participants for each each stimulus and item as follows.

For affect scales, the consensus was the average across all participant ratings.

For the label item, we chose a \textbf{most salient label} as the one most often voted for the respective stimulus. All labels from the starting pool counted as their own \emph{voting label}. However, as participants often added similar, but not completely identical labels, we accumulated the number of newly added labels for each stimulus to represent a single \emph{voting label}. If this accumulated count was chosen more often than starting pool labels, we chose the one most participants entered identically (including correction of spelling mistakes).

\subsubsection{Measurement of association between label and affect ratings}
\label{sec:voctax_method_label_vs_emotion_association}

We employed two methods for measuring of the association between label and affect ratings:

\begin{itemize}
\item \textbf{Predictability of most salient label:} we tested whether aggregated affect ratings allow precise prediction of the most salient label. We modeled the conditional probability $p( \text{label}\ |\ \text{emotion})$ through a multinominal logistic regression model and assessed the goodness of fit through the unweighted average recall metric (alias balanced accuracy) to account for class imbalance. In this context, the metric indicates the degree of overlap between labels regarding their affect ratings: High UAR values indicate that each affect rating is associated with a single label, while low UAR values indicate that each affect rating is associated with multiple labels.

\item \textbf{Correlation of distance matrices:} we tested the correlation of stimuli distances between label-based distances and emotion-based distances. We determined distance matrices as explained in section \ref{sec:voctax_method_distance_calculation} to calculate the Pearson correlations between them. High correlations indicate that stimuli which were rated as similar according to label ratings were also rated as similar according to affect ratings.
\end{itemize}

\section{Results}
\label{sec:voctax_results}

The structure of the results section is as follows: Section \ref{sec:voctax_results_item_reliability} presents the participants perception of the rating items, as well as inter-rater reliability. Section \ref{sec:voctax_results_label_ratings} presents a detailed univariate analysis on the label item ratings, such as derivation of the salient label set. Section \ref{sec:voctax_results_items_association} presents the analysis of associations between rating items, e.g. correlations between affect dimensions as well as label ratings. Section \ref{sec:voctax_results_clustering} performs a cluster analysis on stimuli based on ratings to identify stimuli groups. Finally, section \ref{sec:voctax_results_final_schemes} presents the derived classification schemes which are based on the former results.

\subsection{Rating items reliability and perception}
\label{sec:voctax_results_item_reliability}

Table \ref{tab:voctax_results_item_perception} summarizes (1) raters perception of rating items, and (2) inter-rater reliability (IRR) of rating scales. 

We summarized perception ratings through median and mean values across participants. Ordering rating items ascending by mean perception value, the ranking was valence $\mapsto$ label $\mapsto$ tense arousal $\mapsto$ energetic arousal. We tested the significance of difference in item perception through a pairwise Wilcoxon signed rank test with Bonferroni correction. The difference between valence and energetic arousal ($p < 0.05$) as well as tense arousal ($p < 0.01$) was significant. The difference between valence and label ratings was not significant ($p > 0.05$). The difference between label, energetic arousal and tense arousal was not significant (all $p>0.9$). 

We calculated IRR through intra-class-correlation (ICC) with a two-way random effects model for single raters as recommended by \citet{koo2016guideline} (corresponding to an ICC(2,1) model). Calculation of IRR for label ratings was not possible due to the option of adding custom labels. Ordering rating items descending by mean IRR value, the ranking was valence $\mapsto$ tense arousal $\mapsto$ energetic arousal. \citet{hallgren2012computing} recommends to interpret IRR values as poor for $< 0.4$, fair for $0.4$ -- $0.6$, good for $0.6$ -- $0.75$, and excellent for $> 0.75$. Consequently, reliability was high for valence, fair for tense arousal and poor for energetic arousal. 

\begin{table*}[htbp]
\small
\renewcommand{\arraystretch}{1.3}
\center
\begin{tabular}{@{}l l l l l@{}}
\toprule
                  & \multicolumn{2}{c}{perception} & \multicolumn{2}{c}{IRR (mean (95 \% quantiles))} \\ 
                  & mean         & median          & agreement               & consistency            \\ \midrule
label             & 2.9          & 3 (fair)        & --                       & --                      \\
valence           & 2.2          & 2 (easy)        & 0.70 (0.64 -- 0.76)      & 0.72 (0.66 -- 0.77)     \\
energetic arousal & 3.6          & 3 (fair)        & 0.3 (0.24 -- 0.38)       & 0.35 (0.29 -- 0.43)     \\
tense arousal     & 3.4          & 4 (hard)        & 0.44 (0.37 -- 0.53)      & 0.49 (0.42 -- 0.57)     \\ \bottomrule
\end{tabular}
\caption[Study (A): Summary of rating item perception and reliability]{\textbf{Summary of rating item perception and reliability.} Perception rating range was 1 to 5, with 1 $\hat{=}$ \emph{very easy} and 5 $\hat{=}$ \emph{very hard}. IRR range was -1 to +1, with 0 $\hat{=}$ reliability at chance and 1 $\hat{=}$ perfect reliability.}
\label{tab:voctax_results_item_perception}
\end{table*}

\subsection{Analysis of label ratings}

\label{sec:voctax_results_label_ratings}

\subsubsection{Starting pool synonym pairs}

\SI{43}{\percent} of participants reported \emph{synonym-pairs} for starting labels, i.e. the starting pool to contain at least two synonymous labels. Of those, \SI{90}{\percent} indicated that whimpering and whining where synonymous. We remapped the participant's respective labels to the more frequent one, which was \emph{whining} in all cases. Other synonym pairs were \emph{crying} and \emph{screaming}, \emph{moaning} and \emph{screaming}, and \emph{moaning} and \emph{whining}, all of which were indicated exactly once.

\subsubsection{Newly added labels}

In total, \SI{93}{\percent} of label ratings originated from the starting pool. The quantiles of number of newly added labels per participants were 0 (\SI{0}{\percent}), 0 (\SI{25}{\percent}), 1 (\SI{50}{\percent}), 2.5 (\SI{75}{\percent}), and 12 (\SI{100}{\percent}). Consequently, participants relied mostly on the starting pool labels.

The following labels were added at least twice (after spell checking): 7 $\times$ quietschen (squealing); 3 $\times$ hecheln (panting); 2 $\times$ erz\"ahlen (telling), g\"ahnen (yawning), jauchzen (whooping), singen (singing), spielen (playing).

\begin{figure}[htbp]
    \centering
    \includegraphics[width=0.45\textwidth]{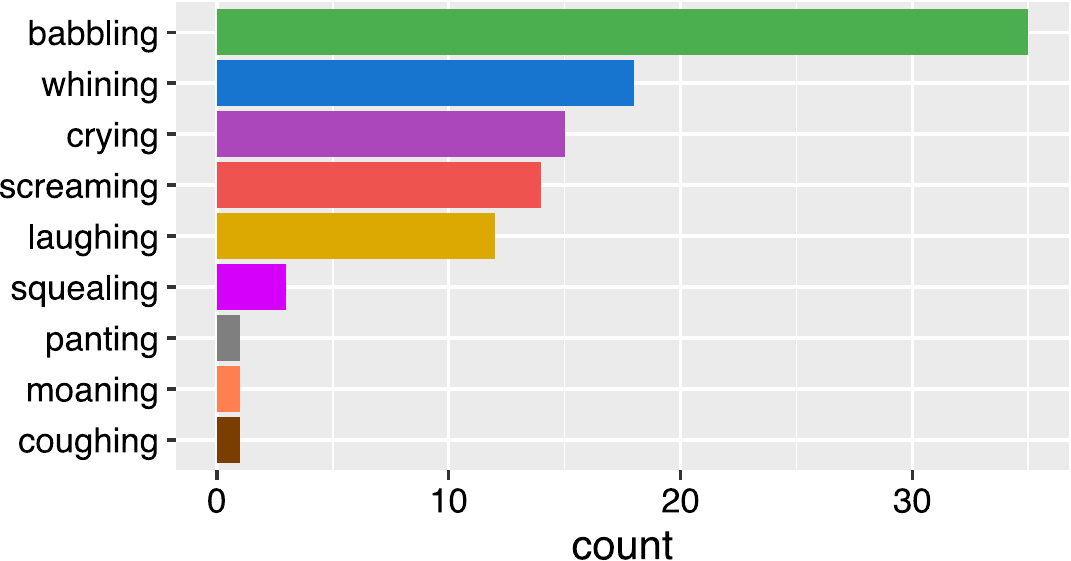}
    \caption[Study (A): Frequency of salient labels]{\textbf{Frequency of salient labels.} Labels are literal English translations from German.}
    
    \label{fig:voctax_results_voclabel_histogram}
\end{figure} 

\subsubsection{Salient label set}
\label{sec:salient_labels}

We determined one consensus label per stimulus as the one most often voted, as the \textbf{salient label} (see section \ref{sec:voctax_method_rating_agreggation}). Figure \ref{fig:voctax_results_voclabel_histogram} shows the frequency of salient labels. This figure, as well remaining in the results section, display literal translations from the original German labels as: schreien = screaming, weinen = crying, jammern = whining, husten = coughing, hecheln = panting, brabbeln = babbling, quietschen = squealing, and lachen = laughing. Two newly added labels \emph{squealing} and \emph{panting} prevailed, while all other labels originated from the starting pool. We highlight that the count of labels does not indicate real-world frequency of the respective call type, but merely the frequency in the data set.

\begin{figure}[htbp]
    \centering
    \includegraphics[width=0.45\textwidth]{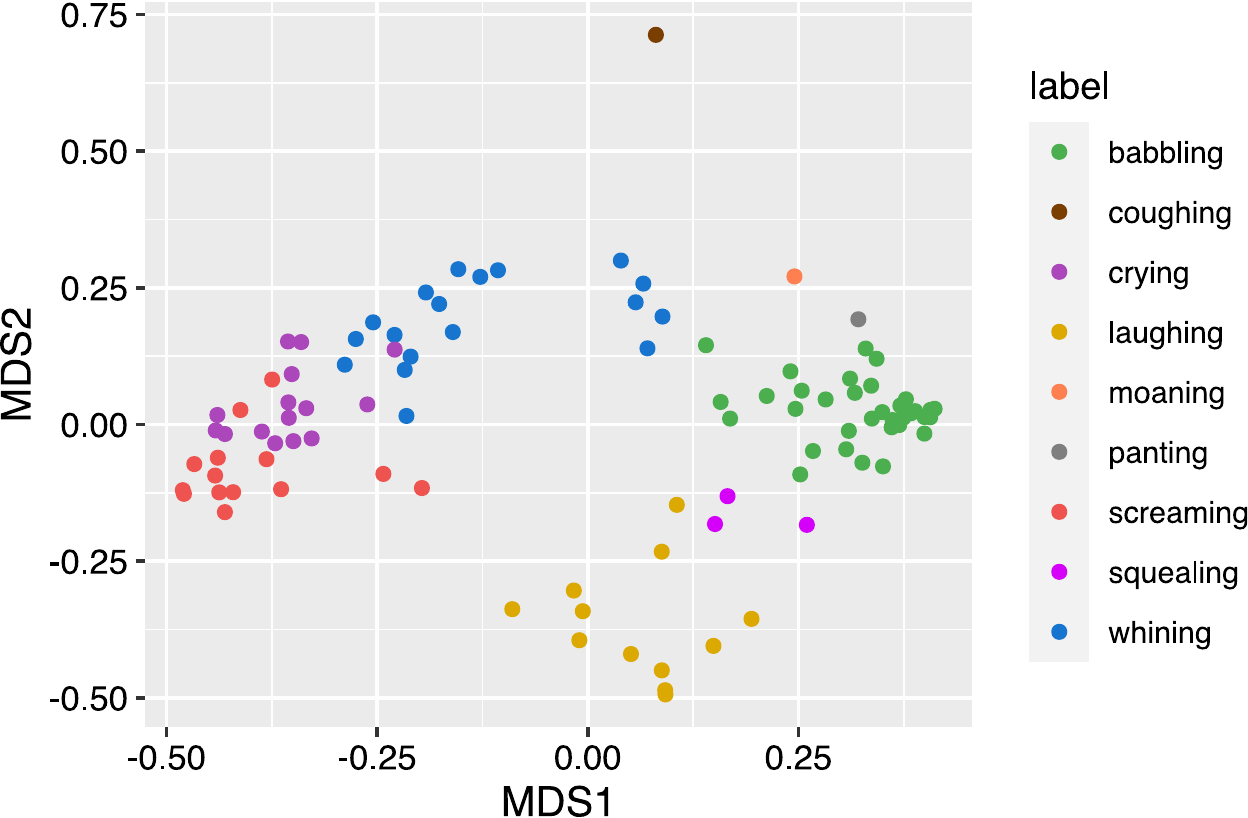}
    \caption[Study (A): Semantic map of salient labels]{\textbf{Semantic map of salient labels}. Data points represent stimuli, color coded by their respective most voted label. Distances between stimuli represent their distances according to label ratings. Technically, this is a non-metric MDS of label-based stimuli distances.}
    
    \label{fig:voctax_results_voclabel_mds}
\end{figure}

Figure \ref{fig:voctax_results_voclabel_mds} shows a semantic map of the vocalization labels. The semantic map was constructed through non-metric multidimensional scaling (MDS) using label-based distances between stimuli (see section \ref{sec:voctax_method_distance_calculation}). MDS is a tool for projecting data points into a $k$ dimensional space so that distances between data points correspond as close as possible to original distances. Unlike a metric MDS, a non-metric MDS interprets distances as rank orders rather than metric distances. The MDS implementation was provided by the R package \texttt{vegan v2.5-7} with function \texttt{metaMDS} with \texttt{monoMDS} engine, global regression mode, and rotation of axis according to the first two principal components. The stress was 0.1. 

We highlight the following observations:

\begin{enumerate}
\item The greatest distance is between crying and babbling vocalizations.  These are placed on opposing ends for the first principal component.
\item The arrangement of stimuli follows a ``tilted horseshoe pattern''. Stimuli are placed around the horseshoe in the following order: screaming $\mapsto$ crying $\mapsto$ whining $\mapsto$ breathing / moaning $\mapsto$ babbling $\mapsto$ squealing $\mapsto$ laughing. Coughing is placed as an outlier far away from all other vocalizations. Laughing and crying represent the open ends of the horseshoe.
\end{enumerate}

\subsection{Association between rating items}
\label{sec:voctax_results_items_association}

\begin{figure*}[htbp]
    \centering
    \includegraphics[width=1\textwidth]{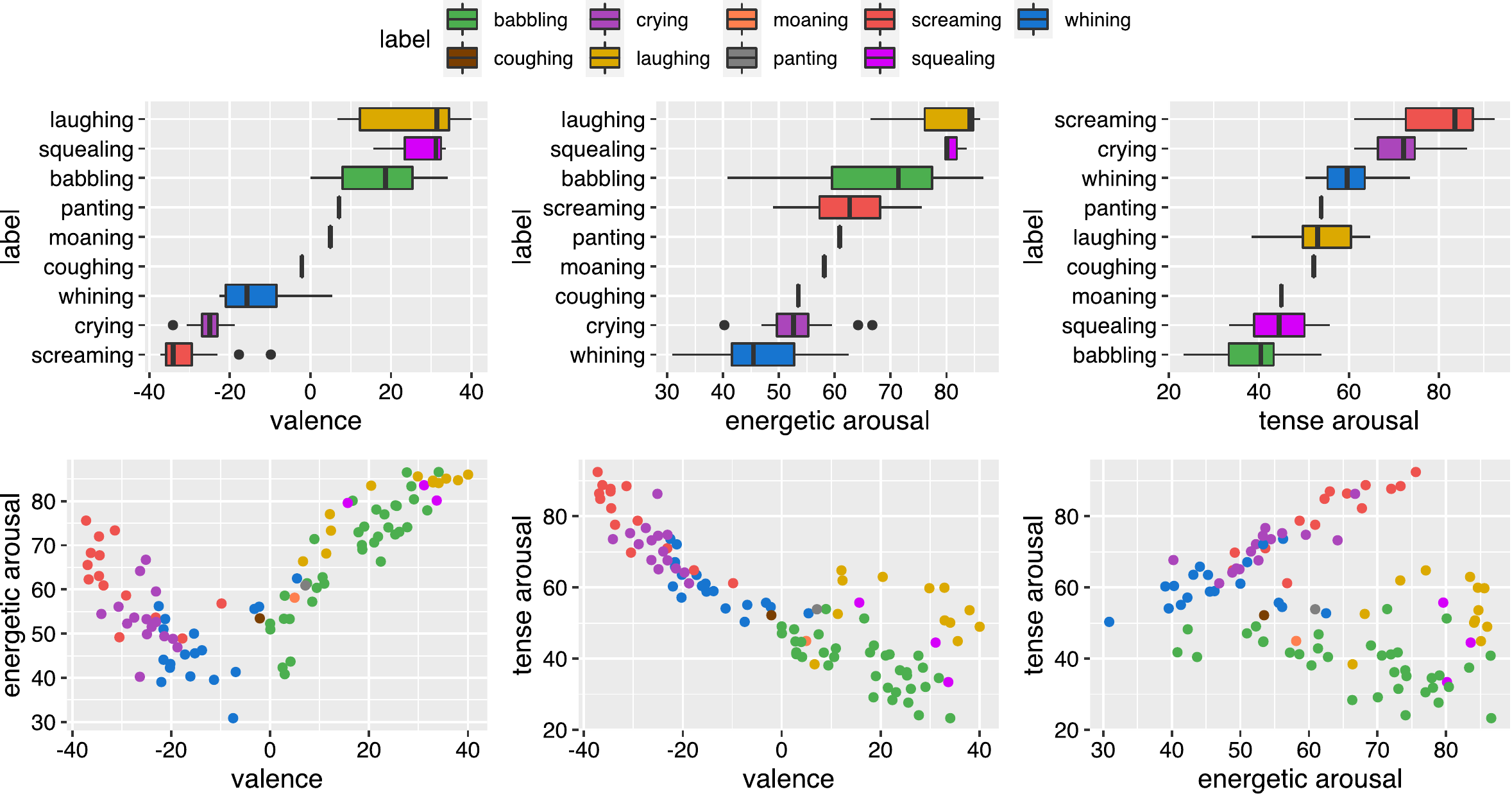}
    \caption[Study (A): Relationship between aggregated label and affect ratings]{\textbf{Relationship between aggregated label and affect ratings.} Each data point represents a stimulus, color coded by its most salient label. A stimulus' affect rating is the average of all participants ratings. The top row shows single affect ratings vs label ratings, bottom row shows dual affect ratings vs label ratings. }
    
    \label{fig:voctax_results_emotions_vs_voclabels}
\end{figure*}

We aggregated participant votes for each stimulus as described in section \ref{sec:voctax_method_rating_agreggation}. Figure \ref{fig:voctax_results_emotions_vs_voclabels} shows the relationship between  aggregated affect and label ratings. 

\subsubsection{Association between affect scales}

We highlight the following observations on the associations between affect scales:

\begin{enumerate}
\item The Pearson correlation between affect scales are: 0.66 for valence \& energetic arousal, -0.84 between valence \& tense arousal, and -0.33 for energetic \& tense arousal. Consequently, valence and tense arousal have the highest association.
\item All affect scales are highly correlated inside the negative valence space. When sub-setting stimuli to those with valence $<0$, Pearson correlations are $-0.57$ for valence \& energetic arousal, $-0.89$ for valence \& tense arousal, and $0.81$ for energetic \& tense arousal.
\item  When excluding stimuli with label \emph{laughing}, the correlation between valence and tense arousal is $-0.93$. Consequently, both affect scales essentially represent the same concept, except for laughing vocalizations.
\end{enumerate}

\subsubsection{Association between label and affect ratings}

Section \ref{sec:voctax_method_label_vs_emotion_association} explained the procedure for calculation of the association between label and affect ratings. We calculated associations for all stimuli as well as for stimuli subsets based on aggregated valence ratings: Negative stimuli (average valence ratings $-50$ to $-10$), neutral stimuli ($-10$ to $10$) and positive stimuli ($10$ to $50$). Table \ref{tab:voctax_results_label_vs_emotion_association} shows the resulting values. 

We highlight the following observations:

\begin{enumerate}

\item Valence has the greatest overall association to label ratings. This is supported by the visualizations in Fig. \ref{fig:voctax_results_emotions_vs_voclabels}, as well as the association values for all stimuli shown in Tab. \ref{tab:voctax_results_label_vs_emotion_association}. Tense arousal has the second strongest and energetic arousal the weakest association.

\item  Screaming, crying and whining are highly associated to valence and tense arousal ratings. They might be viewed as a quantization of the negative valence space. Evidence for this are: (a) the arrangement of labels according to valence ratings correspond to their arrangement in the label-based MDS of Fig. \ref{fig:voctax_results_voclabel_mds}, and (b) the high associations between label and valence / tense arousal ratings in the negative valence space shown in Tab. \ref{tab:voctax_results_label_vs_emotion_association}. However, the association is not perfect, i.e. there is some overlap between salient labels emotion-wise.

\item  Babbling spans the greatest range in all affect dimensions among all labels, particularly for energetic arousal. It occupies the entire valence domain $>0$, i.e. ranges from completely neutral to very positive.

\item  Stimuli with neutral valence ratings have the smallest association to affect ratings, i.e. their separability by affect ratings is low. This applies particularly to \emph{vegetative vocalizations} in the taxonomy of \citet{buder2013acoustic}, comprising \emph{panting}, \emph{coughing} and \emph{moaning}. They received nearly neutral ratings for all affect scales, but have \SI{0}{\percent} overlap in label ratings. The respective association values shown in Tab. \ref{tab:voctax_results_label_vs_emotion_association} support this.

\item Labels of stimuli with positive ratings are best differentiated by tense arousal, mainly because is has the highest association to laughing and babbling. Association values in Tab. \ref{tab:voctax_results_label_vs_emotion_association} support this.

\end{enumerate}

\label{sec:label_vs_emotions}

\begin{table}[htbp]
\renewcommand{\arraystretch}{1.3}
\center
\small
\begin{tabular}{@{}L{1.7cm} L{1.7cm} l l l l@{}}
\toprule
                                      &               & all  & neg. & neut. & pos. \\ \midrule
\multirow{3}{2cm}{aggregated ratings}     & valence & 0.374 & 0.732 & 0.374  & 0.424 \\
                                      & energetic arousal & 0.263 & 0.684 & 0.263  & 0.532 \\
                                      & tense arousal     & 0.346 & 0.735 & 0.346  & 0.622 \\\midrule
\multirow{3}{2cm}{distance correlations} & valence  & 0.644 & 0.547 & 0.119  & 0.563 \\
                                      & energetic arousal & 0.423 & 0.320  & 0.280   & 0.263 \\
                                      & tense arousal     & 0.526 & 0.447 & 0.250   & 0.563 \\ \midrule
\end{tabular}
\caption[Study (A): Association between label and affect ratings]{\textbf{Association between label and affect ratings}. Calculation methods are explained in section \ref{sec:voctax_method_label_vs_emotion_association}. \emph{Aggregated ratings} shows the accuracy when predicting the most salient label based on aggregated affect ratings. Cell values indicate UAR values. \emph{Distance correlation} shows the correlation between the label-based distance matrix and the respective affect distance matrix. Cell values indicate Pearson correlation values. Columns indicate whether all stimuli were used, or limited to those with negative, neutral or positive ratings on average. }
\label{tab:voctax_results_label_vs_emotion_association}
\end{table}

\subsubsection{Including affect ratings into label MDS}

\begin{figure*}[htbp]
    \centering
    \includegraphics[width=1\textwidth]{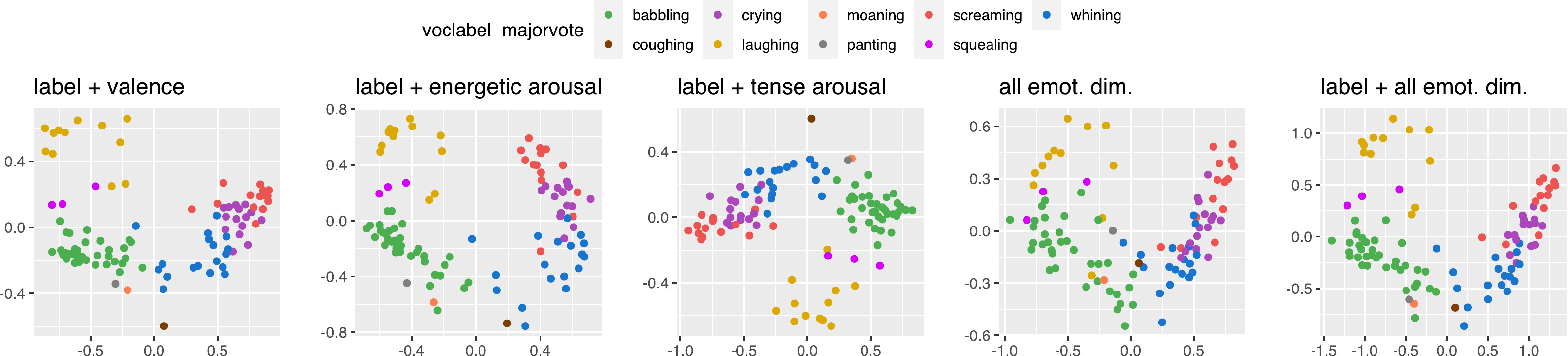}
    \caption[Study (A): Semantic maps for MDS based on combinations of various rating items]{\textbf{Semantic maps for MDS based on combinations of various rating items.} All semantic maps were constructed through MDSs similar to Fig. \ref{fig:voctax_results_voclabel_mds}. Titles of plots indicate the rating items included in the respective MDS.}
    \label{fig:voctax_results_mds_other}
\end{figure*}

We additionally investigated the influence of including distances for affect scales into MDS for producing semantic maps. The goal was to observe changes in stimuli placements and distances. Figure \ref{fig:voctax_results_mds_other} shows the results for various combinations of rating items. We highlight the following observations:

\begin{enumerate}
\item The horseshoe shape observed in the purely label-based MDS (see Fig.  \ref{fig:voctax_results_voclabel_mds}) fundamentally remained even when combining label distances with any or all affect distance matrizes. The central difference to the purely label-based MDS is that neutral reflexive vocalizations \emph{panting}, \emph{coughing} and \emph{moaning} mix into \emph{babbling} and \emph{whining} when affect ratings are considered. As previously explained, this is due to these stimuli being more similar in the affect domain.

\item Even when \emph{excluding} label ratings from the MDS and just using all affect scales a similar horseshoe shape emerges (see plot 4 in Fig. \ref{fig:voctax_results_mds_other}). Again, the difference between \emph{panting}, \emph{coughing}, \emph{moaning} and \emph{babbling} is less pronounced in this case, and the distance between babbling and laughing is less pronounced as well.
\end{enumerate}

\subsection{Clustering of stimuli}
\label{sec:voctax_results_clustering}

In previous sections, we grouped stimuli by their most salient labels. As an alternative approach for stimuli grouping, we performed clustering on label distances. We tested the following clustering algorithms: partitioning around medoids (alias k-medoids) \emph{PAM}, as well as agglomerative clustering variants \emph{average}, \emph{single}, \emph{complete} and \emph{ward} \cite{reynolds2006clustering}. For each algorithm, we evaluated solutions for number of clusters $k \in \{2, \dots, 7  \}$.

For evaluation we used the average silhouette coefficient \cite{rousseeuw1987silhouettes}. This metric indicates the performance of a clustering solutions regarding cluster separability in the value range $[-1, 1]$, where 1 indicates perfect separability between clusters, 0 indicates no structure in data, and -1 indicates worse-than-random separability. We used the implementation of the R package \texttt{cluster v1.2.0} for clustering algorithms and silhouette calculation.

\begin{figure}[htbp]
    \centering
    \includegraphics[width=0.45\textwidth]{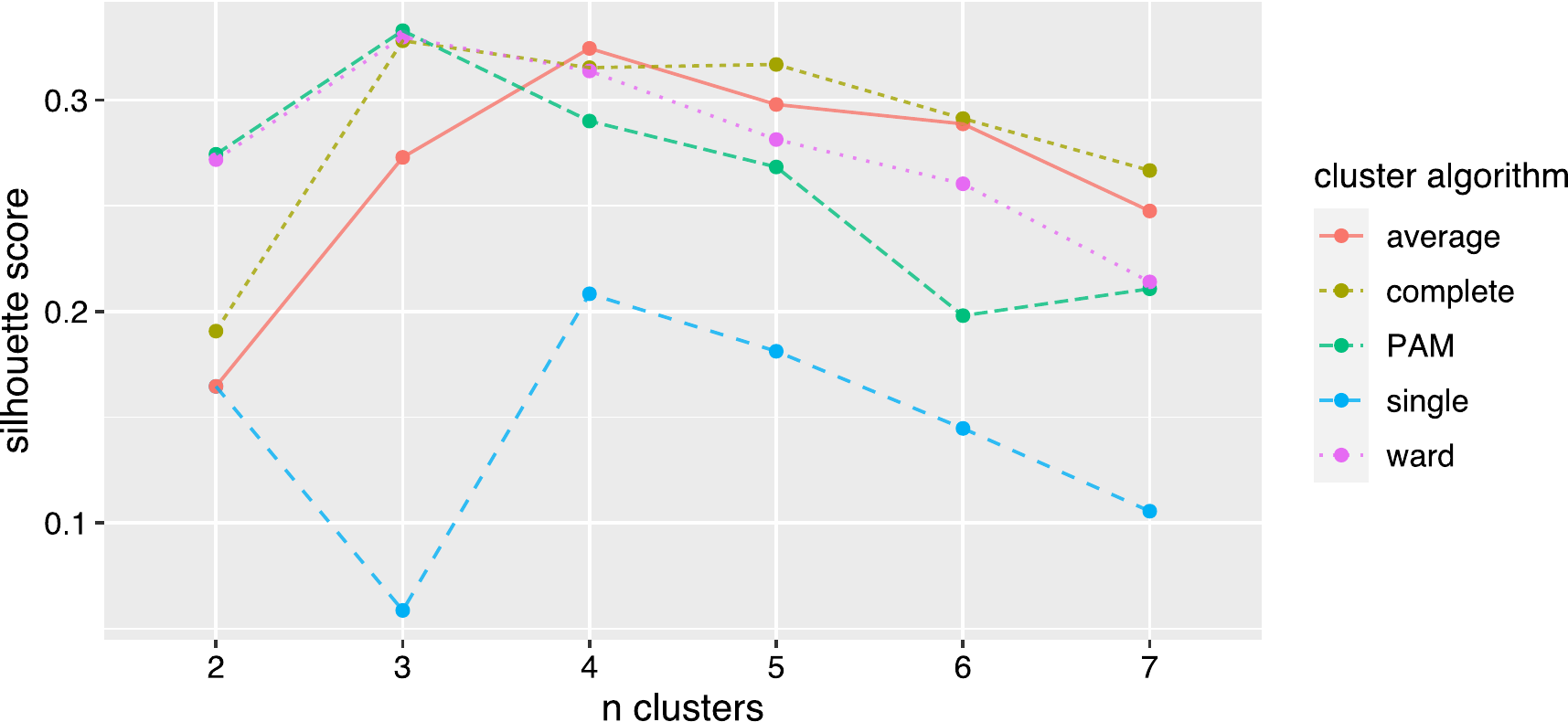}
    \caption[Study (A): Other MDS plots]{\textbf{Average silhouette scores for various clustering algorithms and cluster numbers.} All algorithms are agglomerative clustering variants, except for PAM. }
    
    \label{fig:voctax_results_silscores}
\end{figure}

Figure \ref{fig:voctax_results_silscores} shows the silhouette scores for various cluster numbers and algorithms. We highlight the following observations:
\begin{enumerate}
\item Silhouette scores are generally low, with the highest silhouette score reaching $0.32$ with PAM and 3 clusters. This indicates that stimuli separability based on labels is generally difficult.

\item The best clustering solutions were found for 2,3 and 5 clusters. Silhouette coefficients dropped for more than 5 clusters.
\end{enumerate}

\begin{figure*}[htbp]
    \centering
    \includegraphics[width=1\textwidth]{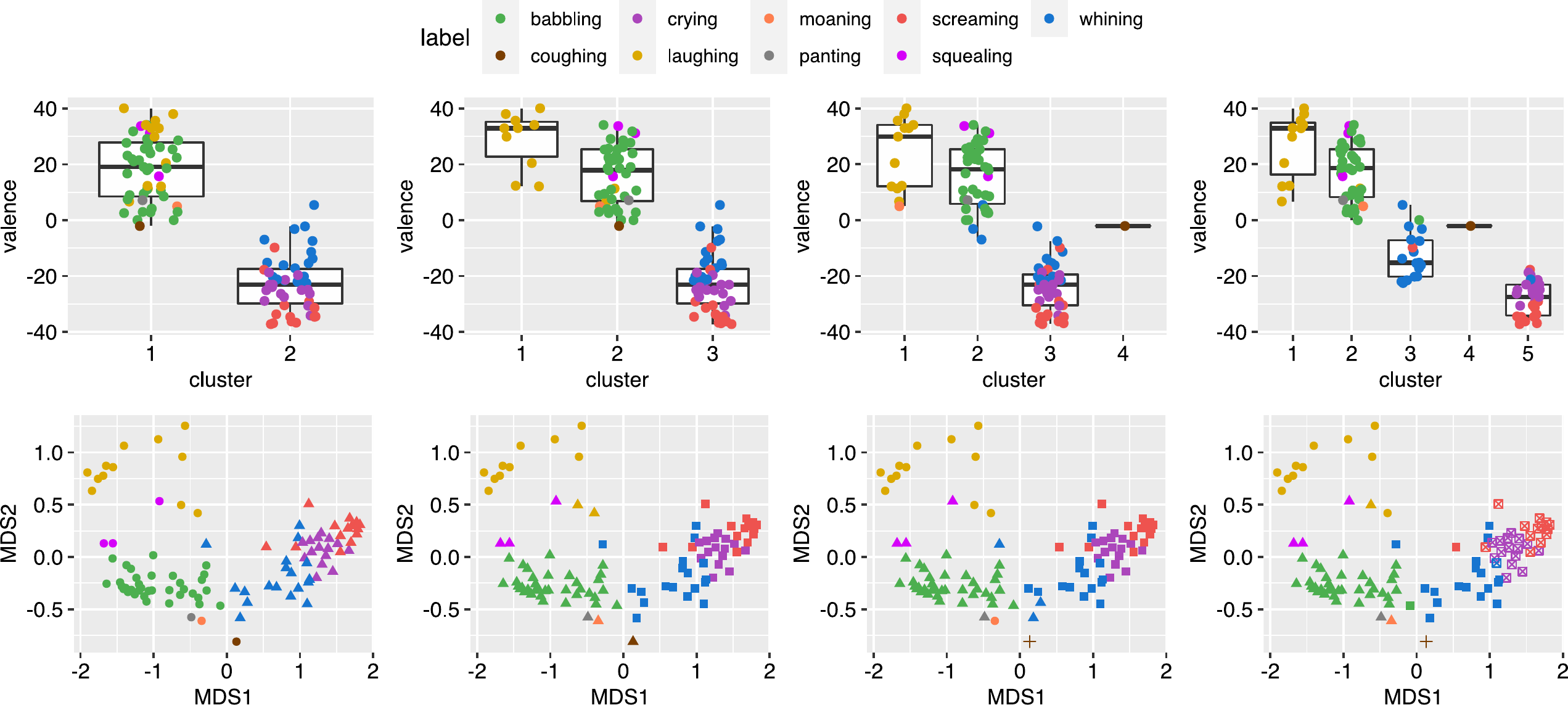}
    \caption[Study (A): Clustering solutions]{\textbf{Clustering solutions.} Each column of plots shows clustering solutions for number of clusters $k=\{2,3,4,5\}$. For each $k$ the clustering was performed by the algorithm which reached the highest silhouette coefficient (see Fig. \ref{fig:voctax_results_silscores}). The top row indicates valence ratings of cluster groups. The bottom row indicates groups through data points shapes in the MDS solutions based on valence + label distances.}
    \label{fig:voctax_results_clustering_solutions}
\end{figure*}

Figure \ref{fig:voctax_results_clustering_solutions} shows the clustering solutions with the highest silhouette score for each number of clusters. $k=2$ differentiates between positive and negative vocalizations. $k=3$ differentiates between laughing, positive non-laughing vocalizations and negative vocalizations. $k=4$ separates coughing, and $k=5$ breaks negative vocalizations into medium negative vocalizations (whining), and medium \& strongly negative vocalizations (crying \& screaming).

\subsection{Derivation of classification schemes}
\label{sec:voctax_results_final_schemes}

\begin{figure*}[thbp]
    \centering
    \includegraphics[width=1\textwidth]{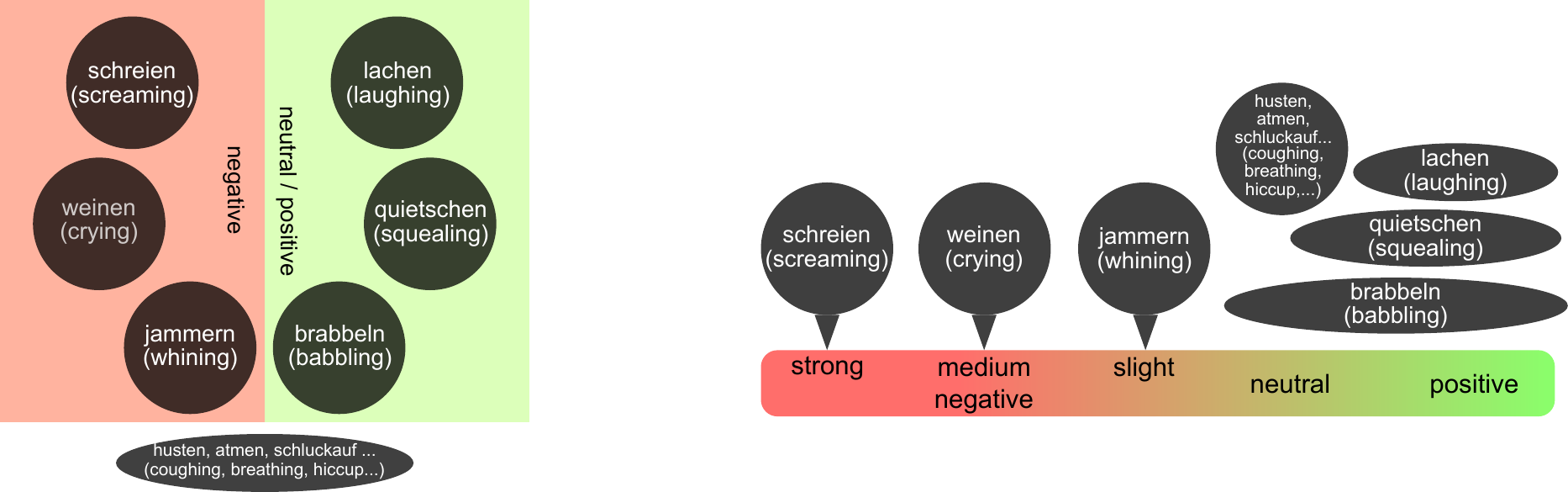}
    \caption[Study (A): Derived classification schemes]{\textbf{Derived classification schemes} Left: Horseshoe scheme. Right: Mood scheme. Circles represent classes. The class class \emph{coughing, breathing, \dots} indicates a group for neutral vegetative vocalizations. Valence related information represents supporting information.}
    
    \label{fig:voctax_results_schemes_both}
\end{figure*}

Based on the previous findings, we proposed two classification schemes shown in Fig. \ref{fig:voctax_results_schemes_both}. Both schemes are stylized semantic maps, i.e. they indicate classes in a spatial arrangement which indicates their relationship to each other \cite{zwarts2010semantic}.

We chose to primarily base these classification schemes on label and valence ratings: Label was the central rating item as it expressed linguistically expressed categories. Affect scales did not completely explain the variability between label ratings as shown in section \ref{sec:label_vs_emotions}, particularly regarding neutral reflexive vocalizations. We chose valence for supporting information as it was the affect scale with the highest inter-rater agreement as well as highest rater perception.

The two schemes are as follows:

\begin{itemize}
\item \textbf{Horseshoe scheme}: Classes are based on the salient labels identified in section \ref{sec:salient_labels}. However, we grouped neutral, voice-less reflexive vocalizations into a single class, comprising sounds such as \emph{coughing, breathing, \dots}. This represents the \emph{vegetative sounds} group in the taxonomy of \citet{buder2013acoustic}. We did this to account for further of such vocalizations such as hiccup, sucking etc., all of which would share the property of involuntary, voice-less vocalizations with neutral affect ratings. The arrangement of classes is a stylized variant of the MDS projections shown in figures \ref{fig:voctax_results_voclabel_mds} and \ref{fig:voctax_results_mds_other}. Class placement particularly highlights similarity between classes based on their acoustic category, for example laughing and crying. The scheme additionally indicates the rough association to valence ranges as supporting information.

\item \textbf{Mood scheme}: Classes correspond to the horseshoe scheme. However, the arrangement of classes indicates their expected valence ranges. Particularly classes screaming, crying and whining are indicated as a quantization of the negative valence space, according to the findings in section  \ref{sec:label_vs_emotions}. Vegetative vocalizations are associated with neutral valence ratings. The remaining classes are merely roughly associated with neutral and positive valence ratings according to their greater valence-wise variance.

\end{itemize}

According to the cluster analysis in section \ref{sec:voctax_results_clustering}, it is to be expected that classes are not perfectly separable. Separability might be increased by combining neighboring classes. For example, the ``resolution'' of the negative vocalizations might be increased by combining them into two or just one group. Also, babbling might be combined with either vegetative vocalizations and or squealing to increase separability.

\section{Discussion}

One of the central results of this study is that \emph{screaming}, \emph{crying} and \emph{fussing} might be viewed as a quantization of the negative valence space. Previous research reflects this: \citet[chapter 2]{barr2000crying} argued that crying is best understood as a graded signal. Pain scales likewise order classes of negative vocalizations as ordinal scales.  

Regarding negative affective vocalizations, participants were likely biased by the starting pool of labels, preventing usage of alternate synonyms. For example, \emph{crying} might have spanned a larger range if \emph{whining} would have not been provided in the starting pool. Evidence for this is that almost all label ratings originated from the starting pool, particularly in the negative valence space. Consequently, future studies might research which labels provide an optimal quantization of the negative valence space, considering  reliability as well as close distress association. 

The final classification scheme shares similarities to the scheme of \citet{buder2013acoustic}. The class \emph{laughing} is present in both; \emph{Crying and fussing} corresponds to \emph{screaming}, \emph{whining} and \emph{fussing}, i.e. our scheme differentiates between three instead of two distress vocalization classes; However, the entire \emph{protophone} category with its numerous subclasses (vowels, grunts, gooing etc.) is contained entirely by the \emph{babbling} class in our scheme. The only protophone subclass separated from \emph{babbling} was \emph{squealing}. This could indicate that laypersons indeed have a low capability to differentiate inside protophone-like vocalizations. However, this could also be an artifact of biasing through the starting pool label set. Possibly, \emph{babbling} was to broad of a term, and participants would have differentiated more if it was absent.

\citet{oller2013functional} introduced a concept named \emph{functional flexibility}: This concepts means protophone vocalizations are flexible regarding the expressed emotion, while reflexive vocalizations are associated with fixed affects. The results of this thesis partially confirm this: Laughing, vegetative vocalizations, and whining/crying/screaming were strictly associated with positive, neutral, and negative valence respectively. However, the \emph{babbling} class, which closely corresponds with the protophone class, spanned the largest valence range, from completely neutral to positive. Consequently, it was indeed associated to a lesser degree with fixed valence values.

Another finding which is supported by our study is that separating vegetative vocalizations from protopohnes usually requires visual confirmation, to determine whether a vocalization was involuntary \cite{oller2013functional,buder2013acoustic}. There were various stimuli for which individual participants provided new labels associated with vegetative classes, such as  \emph{sucking} or  \emph{defication}, but which the great majority labeled as  \emph{babbling}. We hypothesize that  \emph{coughing} and  \emph{breathing} were recognized primarily because they are voice-less.

We highlight that study results might not be directly transferable to English. First and foremost, linguistically expressed categories are inherently influenced by language \cite{anikin2018human}. The literal translations of salient labels into English should be regarded as such: While  \emph{schreien} and  \emph{screaming} are literal translations, \emph{crying} might semantically be the more appropriate term in the context of infant vocalizations, considering the frequency of usage in English research literature \cite{barr1988parental, stark1978classification,buder2013acoustic}. Similarly,  \emph{fussing} is primarily associated with slightly negative vocalizations in these studies, however the literal translation  \emph{aufregen} is not commonly used in German. We hypothesize that  \emph{fussing} corresponds to  \emph{jammern}, which we translated with \emph{whining}. Consequently, it is possible that replication of this study with native English speakers results in the discovery of different classes.

\section*{Acknowledgment}

This work was funded by the research grant for doctoral researchers at Leipzig University of applied sciences, grant number 3100451136, as well as the European Union as part of the ESF-Program, grant number K-7531.20/434-11; SAB-Nr. 100316843.


\appendix

\section{Starting page introductory text}
\label{sec:appendix_voctax_introtext}

\textbf{Stimmung:} Bitte beurteilen Sie, wie ``positiv'' oder ``negativ'' Sie die Simmungslage bzw. die Emotionalit\"at des S\"auglings auf Grundlage der geh\"orten Laut\"au\ss{}erungen einsch\"atzen.
\begin{itemize}
    \item Werte im linken Bereich der Skale bedeuten, dass die Stimmung des S\"auglings eher negativ auf Sie wirkt. Dieser Teil der Skale erstreckt sich von ``sehr negativ'' \"uber ``negativ'' bis ``neutral".
   	\item Werte im rechten Bereich der Skale bedeuten, dass der Stimmung des S\"auglings eher positiv auf Sie wirkt. Dieser Teil der Skale erstreckt sich von ``neutral'' \"uber ``positiv'' bis ``sehr positiv".
\end{itemize}

(<here, participants were presented with a dummy-version of the scale corresponding to the one shown in fig. \ref{fig:voctax_method_survey_mainwindow}>)

Die horizontalen, schwarzen Linien der Skale dienen Ihnen zur Orientierung. Die Skale ist kontinuierlich, das hei\ss{}t, dass alle m\"oglichen Werte zwischen diesen Linien ebenfalls zul\"assig sind. Das selbe gilt f\"ur die n\"achsten beiden Skalen.

\textbf{Wachheit:} Bitte beurteilen Sie, wie hoch Sie den Grad der ``Wachheit'' oder ``Energetisierung'' des S\"auglings auf Grundlage der geh\"orten Laut\"au\ss{}erungen einsch\"atzen.

\begin{itemize}
    \item Werte im linken Bereich der Skale bedeuten, dass der S\"augling eher ``schl\"afrig'' oder ``m\"ude'' oder ``schlapp'' auf Sie wirkt.
    \item Werte im rechten Bereich der Skale bedeuten, dass der S\"augling eher ``wach'' oder ``munter'' oder ``frisch'' auf Sie wirkt.
\end{itemize}      

\textbf{Ruhe:}
Bitte beurteilen Sie, wie hoch Sie den Grad der ``inneren Ruhe'' oder ``Gelassenheit'' des S\"auglings auf Grundlage der geh\"orten Laut\"au\ss{}erung einsch\"atzen.
\begin{itemize}
\item Werte im linken Bereich der Skale bedeuten, dass der S\"augling eher ``ruhig'', ``entspannt'' oder ``gelassen'' auf Sie wirkt.
\item Werte im rechten Bereich der Skale bedeuten, dass der S\"augling eher innerlich ``unruhig'', ``angespannt'', ``aufgeregt'' oder ``nerv\"os'' auf Sie wirkt.
\end{itemize}    

\textbf{Bezeichnung:} Bitte w\"ahlen Sie die Bezeichnung aus, die Ihrer Meinung nach die geh\"orte Laut\"au\ss{}erung am ehesten beschreibt. Beachten Sie dabei die folgenden Punkte:
\begin{itemize}
    \item Sie k\"onnen nur eine Bezeichnung je Aufnahme ausw\"ahlen. Falls Sie eine Aufnahme h\"oren, bei der Sie mehrere Bezeichnungen als zutreffend erachten, entscheiden Sie sich bitte f\"ur diejenige, die f\"ur Sie ``h\"aufiger zu h\"oren'' oder ``pr\"asenter'' ist.
    \item Es ist nicht erforderlich, dass Sie alle Begriffe mindestens einmal im Verlauf der Umfrage verwenden. Wenn Sie zum Beispiel denken sollten, dass zwei der gegebenen Bezeichnungen f\"ur Sie Synonyme darstellen, k\"onnen Sie im Verlauf der Umfrage auch nur eine dieser beiden Bezeichnungen verwenden.
\end{itemize} 

Die gegebene Liste ist nur als Ausgangspunkt zu betrachten. Sie k\"onnen die Liste im Velauf der Umfrage erweitern:
\begin{itemize}
        \item Wenn Sie der Meinung sind, dass keine der bereits gegebenen Bezeichnungen zutrifft, versuchen Sie, eine
        eigene Bezeichnung zu finden. Diese k\"onnen Sie mit Hilfe des Punktes ``Neue Bezeichnung'' hinzuf\"ugen.
        Hinzugef\"ugte Bezeichnungen stehen Ihnen f\"ur die restliche Umfrage dauerhaft zur Verf\"ugung.
        \item  Wenn Sie eine neue Bezeichnung hinzuf\"ugen, muss es sich dabei um ein Verb handeln, dass die
        Laut\"au\ss{}erungen mit einem Wort bezeichnet. Nicht zul\"assig sind beispielsweise zwei Worte, wie ``sich aufregen'',
        oder nicht-Verben, wie z.b. ``\"angstlich".
        \item  Falls Sie keine der gegebenen Bezeichnungen als zutreffend empfinden und Ihnen auch keine eigene,
        passende Bezeichnung einf\"allt, w\"ahlen Sie bitte die Option ``Nicht klar bezeichenbar".
\end{itemize}

\end{document}